# Análisis de la revista cubana *Bibliotecas. Anales de Investigación*

# Analysis of the Cuban journal *Bibliotecas. Anales de Investigación*

**Carlos Luis González-Valiente**
**Sonia Núñez Amaro**
**Javier Ramón Santovenia Díaz**
**Manuel Paulino Linares Herrera**

## RESUMEN

**Objetivo**: Se describe el impacto académico, la calidad del proceso editorial, y las estrategias editoriales y de visibilidad de la revista científica cubana Bibliotecas. Anales de Investigación (BAI), editada por la Biblioteca Nacional de Cuba José Martí.
**Metodología**: Es un estudio descriptivo, con una perspectiva cualitativa a partir de datos cuantitativos que toma como objeto a la revista BAI, de la cual se caracteriza su impacto, gestión editorial y visibilidad. Relativo al análisis de las citas recibidas se utiliza como fuente de consulta la base datos Google Scholar y como herramienta de procesamiento el software EndNote X3. Los indicadores bibliométricos aplicados son: citas por año, citación vs. autocitación, revistas citables vs. documentos no citables, Índice Hirsch y factor de impacto. La calidad del proceso editorial se determina a partir de una autoevaluación efectuada con la metodología de los sistemas de indización SciELO, Scopus, CLASE, Redalyc, Latindex, Dialnet, y ERIH PLUS.
**Resultados**: Se denota una línea de crecimiento en la citación, destacándose como fuentes citantes las revistas científicas de las áreas Bibliotecología y la Ciencia de la Información, Medicina y Ciencias de la Salud, y Educación. Se evidencia que algunos aspectos de formato y contenido inciden negativamente en la calidad del proceso editorial. Se proponen estrategias para mejorar la visibilidad científica a través de la inclusión en bases de datos, directorios, y redes sociales y académicas.
**Conclusiones**: Los resultados obtenidos contribuyen a la toma de decisiones editoriales para aumentar el impacto y la visibilidad científica de BAI.

## ABSTRACT

**Objective**: It is described the academic impact, the quality of the editorial process, and the editorial strategies and of visibility of the scientific Cuban journal Bibliotecas. Anales de investigación (BAI), published by the National Library of Cuba José Martí.
**Methodology**: It is a descriptive study with a qualitative perspective starting from quantitative data that it takes as research object the journal BAI, out of which is characterized its impact, editorial management and visibility. Related to the analysis of received citations, it is used the Google Scholar database as consultation source, and the EndNote X3 software as data processing tool. The bibliometric indicators applied are: citation per year, citation vs. self-citation, citable journals vs. non-citable documents, Hirsch Index and the impact factor. The quality of the editorial process was determined starting from a self-evaluation made with the methodology used by the indexing systems SciELO, Scopus, CLASE, Redalyc, Latindex, Dialnet, and ERIH PLUS.
**Results**: A line of growth is denoted in the citation, standing out as major citing sources the scientific journals from areas like the Library & Information Science, the Medicine and Health Sciences, and the Education. It is evidenced that some aspects related format and content have negatively influenced in the quality of the editorial process. Some strategies are proposed to improve the scientific visibility through the inclusion in databases, directories, and social and academic networks.
**Conclusions**: The obtained results contribute to take editorial decisions for increasing the impact and scientific visibility of BAI.

## INTRODUCCIÓN

El análisis del impacto, la gestión editorial y la visibilidad científica de revistas del área de la Bibliotecología y Ciencias de la Información (BCI) ha sido objeto de análisis dentro del dominio regional iberoamericano (Herrero-Solana & Liberatore, 2008). De manera específica muchos títulos han sido caracterizadas y evaluados, como es el caso de Perspectivas em Gestão & Conhecimento (Ferreira et al., 2014), Ciência da Informação (Pinto, Rodríguez Barquín & Moreiro González, 2006), Anales de Documentación (Santillán-Rivero &Valles-Valenzuela, 2005; González et al., 2008), Revista Interamericana de Bibliotecología (Chinchilla, 2000; Restrepo, 2004), Revista General de Información y Documentación (López et al., 2001), Información, Cultura y Sociedad (Parada, 2009), entre otras.

En el caso particular de Cuba, se disponen de tres revistas orientadas al campo informacional; éstas son: Ciencias de la Información, Revista Cubana de Información en Ciencias de la Salud[1] y Bibliotecas. Anales de Investigación (BAI). Al respecto Mesa Fleitas *et al.* (2006) exploraron la visibilidad de Ciencias de la Información durante el período 1991-2004, mediante indicadores de edición, presentación, productividad y visibilidad; a través de la metodología Agrouniv para la evaluación de producciones científicas. Mientras que Arencibia-Jorge (2008) proyectó su análisis para el caso de la Revista Cubana de Información en Ciencias de la Salud, donde no solo identificó los trabajos más citados sino que también comparó los niveles de impacto de esta revista con los de otras prominentes de América Latina, igualmente orientadas a la BCI.

[1] Conocida anteriormente por el nombre de Acimed.

Sin embargo la revista BAI, a pesar de que ha sido fuente de exploración bibliométrica para indagaciones como las de Pérez Matos (2011), Martínez Rodríguez & Solís Cabrera (2013), González-Valiente, (2014a), González-Valiente (2014b) y González-Valiente (2015); no se conocen sus patrones de impacto, gestión editorial y visibilidad científica dentro de la producción científica bibliotecológica-informativa de Cuba. Es por ello que este estudio tiene como objetivos los siguientes:

1. Describir el impacto académico, la calidad del proceso editorial, y las estrategias editoriales y de visibilidad científica de la revista BAI.
    1.1. Efectuar un análisis de las citas recibidas para valorar el impacto académico alcanzado.
    1.2. Aplicar diagnóstico para evaluar la calidad del proceso editorial.
    1.3. Detallar, a partir de los resultados antes obtenidos, los cambios y las estrategias editoriales orientadas a ganar una mayor visibilidad científica.

## Descripción de la revista *Bibliotecas. Anales de Investigación*[2]

BAI es una revista científica cubana certificada por el Ministerio de Ciencia Tecnología y Medio Ambiente (CITMA) que surgió en 1963, bajo la responsabilidad de la Biblioteca Nacional de Cuba José Martí (BNCJM). Su alcance temático comprende el dominio genérico de las Ciencias de la Información, integrando en ella las tres disciplinas informativo-documentales Archivística, Bibliotecología y Ciencia de la Información; así como toda área que explore, con un carácter interdisciplinar, cualquier fenómeno informacional.

Es reconocida como la revista más antigua del campo informacional en Cuba y América Latina. Sus principales objetivos van encaminados a difundir el quehacer científico del Sistema de Bibliotecas Públicas cubano; así como reflexionar sobre elementos teóricos, prácticos y académicos de la profesión bibliotecaria y del profesional de la información desde una perspectiva científica.

Desde un inicio se concibió como un boletín nombrado *Bibliotecas*, cuya frecuencia durante el periodo 1963-1978 fue bimensual, de 1979 a 1989 semestral, y desde 1992 en lo adelante anual. Posterior al 2002 la revista le incorporó a su título *Anales de Investigación*, incrementando además el rigor científico de los trabajos. Las secciones destinadas a la comunicación científica transitan por un proceso de revisión por pares a doble ciego. Estas secciones son: *artículos científicos* y *reflexiones*; mientras que *experiencias para divulgar*, *reseñas* y *vida científica y académica* son aprobadas a consideración del comité científico. Cada número se publica en formato impreso y electrónico bajo una política de acceso abierto.

---

[2] La información descriptiva provista en esta sección ha sido extraída del portal EcuRed (http://www.ecured.cu/index.php/Anales_de_Investigaci%C3%B3n) y del sitio web oficial de BAI (http://revistas.bnjm.cu/index.php/anales/index).

El idioma de redacción aceptado es el español, el inglés, el portugués, el francés y el italiano. La visibilidad de los artículos siempre ha estado a cargo de sistemas de indización y de registro como las base de datos Cubaciencias y EBSCO, el directorio Latindex, y el índice MIAR. Actualmente conserva en línea todos los números publicados desde 1999.

## METODOLOGÍA

Esta indagación es de tipo descriptiva, con una perspectiva cualitativa que mayormente se basa en el análisis de datos cuantitativos. Se tomó como objeto de estudio a la revista BAI, de la cual se caracterizaron cuestiones como su impacto, gestión editorial y visibilidad científica.

En un primer momento se efectúo un análisis de las citas recibidas, ello con la finalidad de proyectar estrategias editoriales a partir del impacto causado en la comunidad académica. Para este caso se utilizó como fuente de consulta la base datos Google Scholar, en la cual se declararon los términos *Bibliotecas. Anales de Investigación/Bibliotecas*, para el campo *Mostrar artículos publicados en*, con una cobertura de años abierta. Las citas recibidas para los artículos recuperados fueron exportadas para el gestor bibliográfico EndNote X3. En este software se normalizaron las entradas de las referencias para el procesamiento cuantitativo de cada documento, eliminándose los registros duplicados y comprobándose la veracidad de cada dato.

Las principales variables de la investigación que se tuvieron en cuenta fueron: los años de las citas recibidas, tipo y título de las fuentes citantes, número de citas, y las referencias completas de los artículos citados. De estas variables quedaron definidos los siguientes indicadores[3] a calcular:

- *Citas por año*. Número de citas recibidas según los años en los que éstas se han efectuado.
- *Citación vs. autocitación*. Evolución de las citas recibidas contra el número de autocitas que la propia revista se hace durante el periodo comprendido.
- *Revistas citables vs. documentos no citables*. Se consideran documentos citables aquellos que han sido publicados por revistas científicas, de ahí que este indicador muestra la distribución del número total de citas recibidas por revistas (fuentes primarias) contra las citas recibidas desde otras fuentes de carácter secundario.
- *Índice H o Índice Hirsch*. Su cálculo es puramente automático y es realizado por Google Scholar, base de datos que al respecto declara: "el Índice H es el mayor número H, de forma que H publicaciones se han citado al menos H veces. La segunda columna muestra la versión actual del cálculo, que es el mayor número H, de forma que H publicaciones se

3 Algunos de estos indicadores se han adoptado de SCImago Journal and Country Rank (http://www.scimagojr.com/index.php), sitio especializado en el análisis de revistas científicas indizadas por SCOPUS.

han citado al menos H veces más en los últimos 5 años" (Google Scholar, 2014).

- *Factor de impacto*. Calculado a partir de las citas recibidas en el 2015, según Google Scholar, para los dos últimos años (2013-2014). Las autocitas no se consideraron y la fórmula aplicada fue la definida por Garfield (Spinak, 1996, p. 126):

$$\text{Impacto (i)} = \frac{CIT_{i-1}\,(i) + CIT_{i-2}\,(i)}{PUB_{i-1}\,(i) + PUB_{i-2}\,(i)}, \text{ donde:}$$

i: es el año corriente.
$CIT_{i-1}$: citaciones recibidas en el año corriente a artículos publicados en el año (i-1), análogo para (i-2).
$PUB_{i-1}$: cantidad de artículos publicados en el año (i-1), análogo para (i-2).

La medición de los datos se hizo a través del conteo, con el apoyo del programa Microsoft Excel, el cual también se utilizó para la generación de gráficos y figuras.

Los resultados obtenidos desde el análisis de citación constituyeron un punto de partida para la proyección futura y las estrategias editoriales de la revista. Este aspecto se solidificó con la ejecución de una autoevaluación de los números publicados desde el 2009 hasta el 2014; a partir de metodologías empleadas por los sistemas de indización SciELO, SCOPUS, CLASE, Redalyc, Latindex, Dialnet, y ERIH PLUS. Esto contribuyó a identificar la calidad actual del proceso editorial.

## RESULTADOS Y DISCUSIÓN

### *Análisis del impacto de Bibliotecas. Anales de Investigación*

Se han identificado 85 citas para el periodo 1996-2016. Véase a través de la figura 1 que durante los primeros diez años (1996-2006) las citaciones eran muy bajas, para un promedio de 1,6 por año. Además la irregularidad en la citación era alta, teniendo en cuenta que en los años 1997, 1998 y 1999 ningún trabajo fue citado. Ya a partir del 2007 este promedio aumentó a un 6,9 por año, destacándose el 2014 como el de mayor impacto. En este año los trabajos citantes han sido generalmente producidos por autores cubanos y publicados por revistas nacionales del área de la BCI, la Educación, y las Ciencias de la Salud.

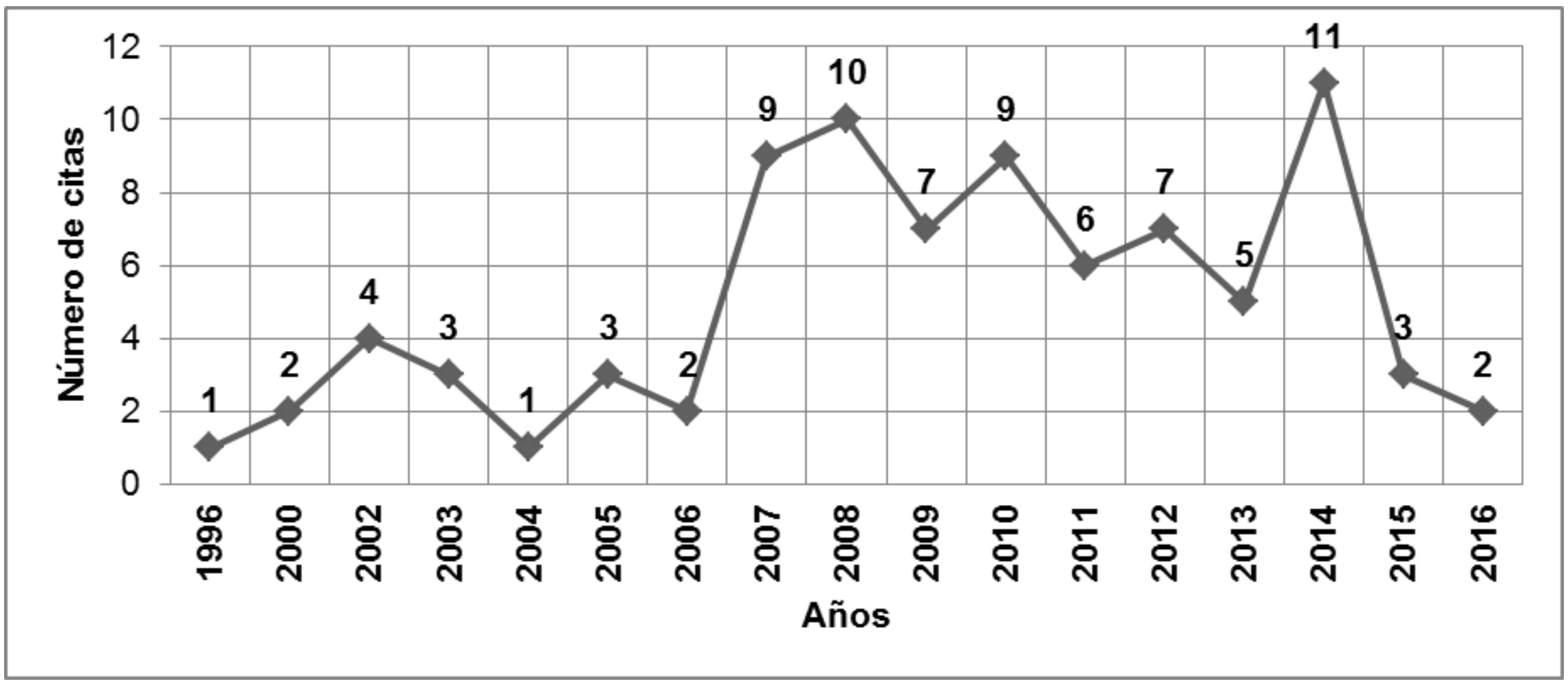

**Figura 1**. Evolución de las citas por año (*Fuente: los autores*, 2015).

Tras una correlación de la autocitación contra las citas provenientes de fuentes externas, se distingue que el promedio de autocitación es ligeramente ínfimo, para un 1,11%, el cual presenta un cierto nivel de decrecimiento según la secuencia en años (véase figura 2). Estas autocitaciones solo representan un 22,35% del total de citas identificadas, tendencia considerada como positiva ya que esto demuestra la marcada visibilidad e impacto que la revista muestra para fuentes externas.

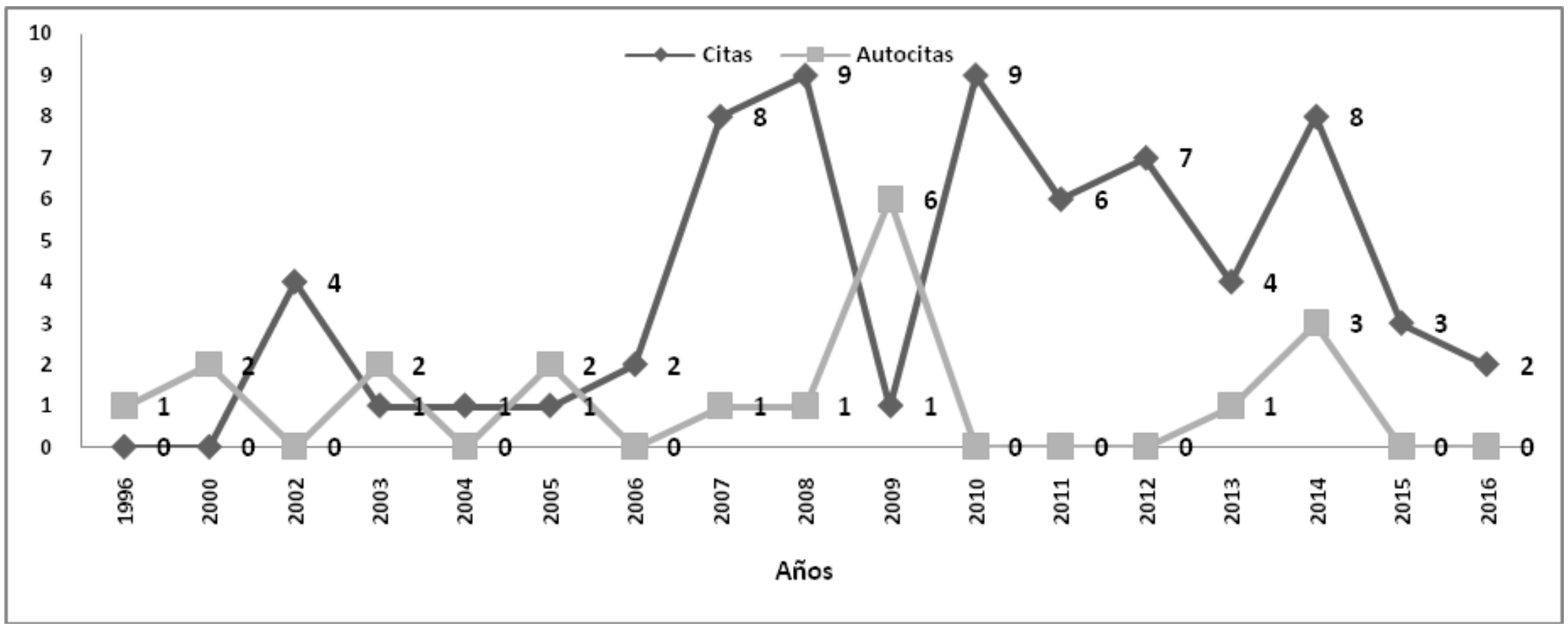

**Figura 2.** Evolución de la citación *vs*. autocitación (*Fuente: Los autores*, 2015).

La mayoría de estas citas son efectuadas por revistas científicas (72,94%), figurando en menor medida las tesis (11,76%), los artículos electrónicos no revisados por pares (8,23%), las ponencias de congreso (4,7%) y los libros (2,35%). Nótese a través de la figura 3 que las citaciones por revistas alcanzaron su mayor prominencia y regularidad durante el periodo 2007-2010, con cierto nivel de incremento. Indistintamente se percibe que el ritmo de citación es más lineal para los documentos no citables.

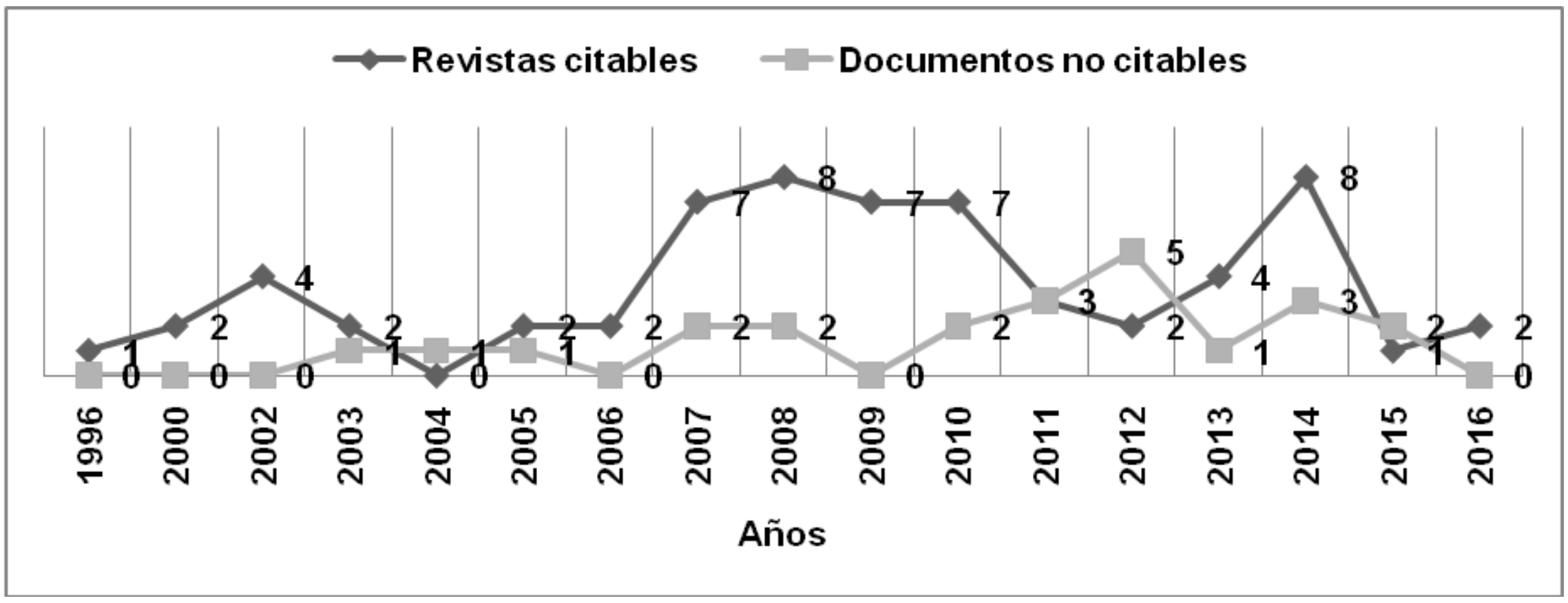

**Figura 3.** Comportamiento de la citación de las revistas científicas *vs.* documentos no citables (*Fuente: Los autores*, 2015).

Haciendo énfasis en las revistas citantes, éstas como literatura científica primaria, se denota que las extranjeras (51,72%) superan a las nacionales (48,27%), para un total de 29 títulos. De las extranjeras se destacan las procedentes de la propia región latinoamericana (24,13%) y en menor medida las europeas (20,68%) (Véase tabla 1). Esto puede que devenga gracias al español como idioma principal de publicación de BAI, algo que ha incidido sobre los trabajos citantes ya que el 94,13% se encuentran en este idioma, mientras que el resto solo está en inglés (2,35%), portugués (2,35%) y francés (1,17%).

Precisamente de los títulos cubanos citantes se destacan aquellos orientados a la producción científica sobre BCI, tales como: BAI, Revista Cubana de Información en Ciencias de la Salud, y Ciencias de la Información (Véase tabla 1). Se destacan revistas de áreas temáticas como la BCI (41,37%), la Medicina y las Ciencias de la Salud (31,03%), y la Educación (17,24%). Es importante resaltar que del total de revistas citantes un 48,27% es de la corriente principal, ya que actualmente se encuentran indizadas por bases de datos como Web of Science o Scopus. Al respecto se detectan títulos de gran impacto internacional como *Scientometrics*, *The International Information & Library Review*, *Information Research*, e *International Biodeterioration & Biodegradation*.

| **Revista** | **País** | **Citas** |
|---|---|---|
| Bibliotecas. Anales de Investigación | Cuba | 19 |
| Revista Cubana de Información en Ciencias de la Salud | Cuba | 13 |
| Ciencias de la Información | Cuba | 3 |
| Investigación Bibliotecológica | México | 2 |
| Anuario ThinkEPI | España | 1 |
| Biblios | Perú | 1 |
| Cuadernos de Educación y Desarrollo | España | 1 |
| Educación Médica Superior | Cuba | 1 |
| Infociencia | Cuba | 1 |
| Information Research | Inglaterra | 1 |
| International Biodeterioration & Biodegradation | Inglaterra | 1 |
| Revista de la Facultad Nacional de Salud Pública | Colombia | 1 |
| Revista Museo de La Plata, Sección Botánica (Argentina) | Argentina | 1 |
| Revista de la Sociedad Venezolana de Microbiología | Venezuela | 1 |
| Revista CENIC. Ciencias Biológicas | Cuba | 1 |
| Revista Cubana de Estomatología | Cuba | 1 |
| Revista Cubana de Higiene y Epidemiología | Cuba | 1 |
| Revista Cubana de Informática Médica | Cuba | 1 |
| Revista Cubana de Salud y Trabajo | Cuba | 1 |
| Revista de la Biblioteca Nacional de Cuba José Martí | Cuba | 1 |
| Revista EDUSOL | Cuba | 1 |
| Revista Electrónica Conrado | Cuba | 1 |
| Revista Interamericana de Bibliotecología | Colombia | 1 |
| Scientometrics | Países Bajos | 1 |
| Scire | España | 1 |
| Simbiosis | Puerto Rico | 1 |
| The International Information & Library Review | E.U.A. | 1 |
| Transformación | Cuba | 1 |
| Universidad y Sociedad | Cuba | 1 |

**Tabla 1.** Citas recibidas según el tipo de referencia (*Fuente: Los autores, 2015*).

Las 85 citas identificadas se refrieren a 39 documentos, de los cuales solo 16 han resultado cocitados. En la tabla 2 se presentan las referencias de los artículos que han recibido al menos dos citas, siendo muy distintivo que los más citados se han publicado antes del 2000. Por otro lado han sido de gran impacto los artículos que aboran temáticas referidas al desarrollo bibliotecológico-informativo cubano (ej.: Setién Quesada, Llorente & García, 1990; Setién Quesada, 1995; Setién Quesada, 2005), los estudios cuantitativos y cualitativos de la ciencia y la tecnología (Pérez Matos, 1998; Pérez Matos, 2003; González, 2003; Peralta & Frías, 2011), la alfabetización informacional (Cárdenas & Jiménez, 2007; Quindemil Torrijos, 2008; González-Valiente, Sánchez-Rodríguez & Lezcano-Pérez, 2013), la promoción de la lectura (Setién Quesada, 1991; González, 2003; Rivera & Lazcano, 2003; Castillo & Martí, 2006), una temática convertida en líder en los años 90 debido a su importante rol dentro del trabajo bibliotecario; los procesos técnicos bibliotecarios (Escobar, 1998; Collazo, 2003; Hernández Concepción, 2005; Libera, Machado & Núñez, 2005; Hidalgo & Borrego, 2006), entre otras. Muchas de estas investigaciones han sido impulsadas por el Departamento de Investigaciones Histórico-Cultural y Bibliotecológico de la BNCJM, ejemplo de ello es el proyecto de investigación sobre la bibliografía cubana iniciado en 1996, del cual emergieron gran parte de los estudios bibliométricos publicados.

| **Autor(es)** | **Título y fuente** | **Año** | **# de citas** |
|---|---|---|---|
| Emilio Setién Quesada | Aportes metodológicos sobre la actividad bibliotecaria en el Ministerio de Cultura de Cuba. (2), 24. | 1995 | 9 |
| Concha González | Los bibliobuses como instrumento de fomento de la lectura. (1-2), 173-190. | 2003 | 7 |
| Yahumila Hidalgo Cerito; Sofía Borrego Alonso | Aislamiento y caracterización de hongos en documentos de la Biblioteca Nacional José Martí. *2*, 95-101. | 2006 | 6 |
| Nuria Esther Pérez Matos | ¿Bibliometría o bibliotecometría? (1-2), 38-61. | 2003 | 5 |
| Emilio Setién Quesada; Llorente, M., & García, V. M. | Bibliotecología cubana y actividad científico-informativa. *28*(2), 37-44. | 1990 | 5 |
| Eneida Quindemil Torrijos | Políticas de información y su incidencia en la alfabetización informacional. Consideraciones desde la perspectiva cubana. *4*, 28-35 | 2008 | 5 |
| Emilio Setién Quesada | Estudios sobre el trabajo con los lectores en las bibliotecas. Su relación con el Programa Nacional de la Lectura. *29*, 48-58. | 1991 | 4 |
| Nuria Esther Pérez Matos | Algunas experiencias del estudio bibliométrico de la bibliografía cubana. *1-2*, 29-47. | 1998 | 3 |
| Editorial | Los estudios bibliotecológicos en Cuba. *10*(11), 49-55. | 1972 | 3 |
| Arianne de Cárdenas Cristiá, N. Jiménez Hernández | Acceso universal a la información: de la ecuación de usuarios a la alfabetización informacional. 3, 5-40 | 2007 | 2 |
| Zoia Rivera, Dayilién Lazcano | Biblioteca pública del Lyceum Lawn Tennis Club: promotora de la cultura en la Cuba republicana. | 2003 | 2 |
| Norma Barrios | La gestión de información y sus recursos (parte I). 80-104. | 2001 | 2 |
| Carlos Luis González-Valiente, Yilianne Sánchez-Rodríguez, | Propuesta de un programa de alfabetización informacional para los estudiantes de la Universidad de la Habana. *8-9*(8-9), 121-131. | 2013 | 2 |
| Yazmín Lezcano-Pérez | | | |
| Editorial | Seminario-taller sobre promoción de la lectura en las bibliotecas. *29*(1-2), 5-23. | 1991 | 2 |
| Emilio Setién Quesada | Raíces y concepciones que conducen a la teoría bibliológico informativa. (1), 35-53. | 2005 | 2 |
| Araceli García Carranza | La bibliografía cubana: inventario de nuestra cultura cubana. *28*(1), 25-51. | 1990 | 2 |

**Tabla 2.** Artículos más citados (*Fuente: Los autores, 2015*).

De los trabajos más citados también se destacan los producidos por la DraC. Nuria Esther Pérez Matos y el Dr.C. Emilio Setién Quesada, dos autores cuya afiliación es la BNCJM y que desde allí han contribuido con la investigación bibliotecológico-informativa en el contexto tanto cubano como latinoamericano.

En general, todas las citaciones que ha recibido BAI hacen que posea un Índice H de 5, un índice relativamente considerable para la poca visibilidad que poseía la revista antiguamente. Mientras que su factor de impacto es de 0.05, de acuerdo con los cálculos realizados mediante la fórmula siguiente:

$$\text{Impacto (2015)} = \frac{2\,(2013) + 0\,(2014)}{17\,(2013) + 20\,(2014)} = 0.05$$

*Evaluación de la calidad editorial de BAI*

Una estrategia para evaluar la calidad editorial de BAI fue la autoaplicación de criterios e indicadores definidos por variados sistemas de indización, tales como bases de datos y directorios regionales e internacionales. Entre los regionales han figurado SciELO, Clase, Redalyc, Latindex y Dialnet; mientras que los internacionales han sido Scopus y ERIH PLUS.

Cada uno de estos sistemas declara diversa cantidad de indicadores, orientados indistintamente a evaluar aspectos de formato, contenido e impacto. Aquí se hará mayor énfasis en las cuestiones de formato y contenido, puesto que el impacto ya fue caracterizado en la sección anterior.

En la tabla 3 se presenta una matriz de los resultados cuantitativos obtenidos para cada año en específico y para el periodo en general que comprende la evaluación, 2009-2014. Los resultados más bajos se obtuvieron desde sistemas como Scopus y Clase, en tanto que el mayor cumplimiento se refiere a lo establecido por ERIH PLUS.

| | | **Cumplimiento de los criterios de evaluación (cumplidos/total)** | | | | | | |
|---|---|---|---|---|---|---|---|---|
| **Años** | **Artículos** | **SciELO** | **SCOPUS** | **CLASE** | **Redalyc** | **Latindex** | **Dialnet** | **ERIH PLUS** |

| | **indizables** | | | | | | | |
|---|---|---|---|---|---|---|---|---|
| 2009 | 10 | 15/18 | 11/14 | 24/28 | 11/12 | 34/36 | 2/3 | 6/6 |
| 2010<br>2011 | 15 | 15/18 | 10/14 | 24/28 | 10/12 | 33/36 | 2/3 | 6/6 |
| 2012<br>2013 | 17 | 15/18 | 10/14 | 24/28 | 10/12 | 33/36 | 2/3 | 6/6 |
| 2014 | 20 | 15/18 | 11/14 | 24/28 | 11/12 | 34/36 | 2/3 | 6/6 |
| **Total** | | 15/18 | 10/14 | 24/28 | 10/12 | 33/36 | 2/3 | 6/6 |

**Tabla 3**. Resultados de la aplicación de indicadores de calidad editorial propuestos por los sistemas de indización (*Fuente: los autores, 2015*).

Es de destacar que aunque los indicadores cumplidos no están muy distantes del total, muchas veces un criterio puede ser motivo sólido y suficiente para que un sistema no apruebe la inclusión de la revista. Desde el punto cualitativo las principales debilidades han sido:

- Los artículos no están identificados mediante un membrete bibliográfico que contenga aspectos como nombre de la revista, ISSN, número, volumen y periodo que cubre la edición.
- La frecuencia anual es muy baja, principalmente para sistemas como SciELO, Scopus y Clase.
- Incumplimiento de la puntualidad declarada en la periodicidad, fundamentalmente para los números del 2010 y 2012.
- No posee registro en el catálogo de Latindex, un elemento indispensable para Dialnet.
- Estructura de las secciones un poco alejadas de lo sugerido por Clase, SciELO y Scopus.
- Alto índice de la autoría local.
- Contenidos con abordajes un tanto locales y de poca proyección internacional.

### *Estado actual y proyecciones futuras*

A partir del 2015 la revista se ha propuesto entrar en una nueva fase superior donde la prioridad es la calidad y visibilidad científica de sus artículos. Con los resultados arrojados en los análisis efectuados anteriormente se comenzaron a producir cambios desde el segundo semestre de 2015. Las principales modificaciones que tuvieron lugar a lo interno de la gestión editorial fueron:

- Reconformación del comité editorial por miembros más comprometidos.

- Lanzamiento de nuevas normas editoriales.
- Rediseño del modelo de arbitraje con vistas a generar mayor rigor en las evaluaciones de los artículos.
- Aumento del número de revisores por artículos con gran equilibrio de árbitros nacionales e internacionales.
- Incorporación del membrete bibliográfico en las cubiertas, al inicio del documento y en cada página del documento.
- Cambio de la frecuencia de publicación de anual a bianual a partir del 2016, con una salida en junio y otra en diciembre.
- Modificación de la estructura de las secciones para la tipología: Artículos originales, Artículos de revisión, Comunicaciones, Estudios de caso, Reseñas y Experiencias para divulgar.
- Utilización del software *Open Journal System* para la publicación de los artículos bajo la licencia *Creative Commons Reconocimiento-No Comercial-Compartir Igual 4.0 Internacional License*, con una disposición en línea en la siguiente dirección: http://revistas.bnjm.cu/index.php/anales/index.
- Incorporación y registro en nuevos sistemas de información científica.

Relativo a la incorporación de la revista en sistemas de indización, ésta estuvo registrada y procesada antiguamente por Cubaciencias, Latindex, MIAR y EBSCO; tal y como se mencionaba en una de las secciones anteriores. Con la finalidad de aumentar la visibilidad científica, a partir del segundo semestre de 2015 se ha garantizado la inclusión de BAI en nuevos sistemas como: E-LIS, Google Scholar, CiteFactor, JournalsTOC, BASE, ERIH PLUS y JIFACTOR (véase tabla 4). En tanto que se está procesando para ser incluida en Redalyc, Clase y Dialnet. También se ha priorizado su ingreso en las redes sociales Facebook, Twitter y un Blog soportado por la plataforma cubana Reflejos de Cubava.cu; así como también en redes sociales académicas como LinkedIn, Academia.edu y Mendeley. Todo ello para establecer espacios de intercambio e interactividad con los lectores, investigadores, bibliotecarios y demás profesionales interesados por los materiales publicados y las notificaciones de la revista. Considerar estas redes sociales, más allá de los habituales sistemas científicos deviene en estrategia indispensable, ya que se ha evidenciado que las revistas con presencia en plataformas como éstas alcanzan un mayor índice de impacto (Herrero-Gutiérrez, Álvarez-Nobell & López-Ornelas, 2011; Oller Alonso, Segarra Saavedra & Plaza Nogueira, 2012; Segado-Boj, 2013).

| Sistema de indización | Descripción | Orientación temática |
|---|---|---|
| **Cubaciencias** | Base de datos desarrollada en 1997 por el Instituto de Información Científica y Tecnológica (IDICT) de Cuba. Procesa publicaciones seriadas, manuscritos, informes, eventos y tesis de doctorado. | Ciencias Técnicas y Aplicadas, Biomedicina, Ciencias Sociales y Ciencias Agropecuarias. |
| | Actualmente cuenta con más de 43 mil registros. | |
| **E-LIS** | Repositorio temático de acceso abierto. | Bibliotecología y Ciencia de la Información |
| **Latindex** | Sistema Regional de Información en Línea para Revistas Científicas de América Latina, el Caribe, España y Portugal. | Multidisciplinar |
| **MIAR** | La Matriz de Información para el Análisis de Revistas es una herramienta colaborativa que ofrece un índice de revistas con el objetivo de incrementar su visibilidad. | Multidisciplinar |
| **EBSCO** | Compañía norteamericana fundada en 1994 perteneciente a la *EBSCO Industries, Inc*. Opera bajo suscripción ofreciendo acceso a bases de datos, libros electrónicos, revistas, etc. | Multidisciplinar |
| **ERIH PLUS** | Índice abierto de revistas conocido como *European Reference Index for the Humanities and Social Sciences*, orientado a las disciplinas humanísticas y de ciancias sociales. | Ciencias Humanas y Ciencias Sociales |
| **Google Scholar** | Buscador académico de Google. | Multidisciplinar |
| **CiteFactor** | Directorio de acceso abierto que provee indización de importantes revistas y actas de eventos. | Multidisciplinar |
| **JIFACTOR** | El *Journal Impact Factor* es un directorio de acceso gratuito que provee herramientas para el posicionamiento, la categorización, la evaluación y la comparación de revistas científicas. | Multidisciplinar |
| **JournalsTOC** | Sistema de alerta gratuito que colecciona metadatos de tabla de contenidos de revistas | Multidisciplinar |
| | provenientes de más de 2673 editoriales. | |
| **BASE** | Motor de búsqueda de acceso abierto conocido como *Bielefeld Academic Search Engine*, operado por la *Bielefeld University Library*. Se encarga de recolectar, normalizar e indizar datos provenientes de más de 3 mil fuentes. | Multidisciplinar |

**Tabla 4.** Sistemas de indización donde está registrada BAI (*Fuente: Los autores, 2015*).

Las proyecciones futuras de BAI están encaminadas a cuestiones de contenido, aspecto que solo impulsa y garantiza la visibilidad científica en sistemas de indización de prestigio y el impacto académico evidenciado a través de indicadores bibliométricos, altmétricos y de citación. Las principales estrategias se dirigen hacia:

- Aceptación de artículos con alto rigor científico y de gran aporte para el campo disciplinar, de acuerdo a las tendencias disciplinares actuales.
- Mayor incidencia en la verificación de la actualidad y calidad de la bibliografía empleada en las investigaciones.
- Mayor apertura hacia la investigación cuyos contenidos no sean tan locales y sí de gran interés para la comunidad científica internacional.
- Publicación de artículos en idioma inglés, éste como idioma mundial para la comunicación científica.
- Priorización de manuscritos en los que predomine la coautoría.
- Cierto equilibrio en la publicación de estudios con autoría y colaboración internacional.
- Proyección hacia nuevas bases de datos.

## CONSIDERACIONES FINALES

El propósito de este estudio ha sido el de mostrar una panorámica de los niveles de citación que ha alcanzado BAI, todo ello con vistas a la toma de decisiones para la mejora de los procesos editoriales y de visibilidad científica. Al no ser considerada esta revista de la corriente principal, Google Scholar se ha convertido en una fuente de óptima consulta para los análisis aquí efectuados, ya que es reconocida como una base de datos bien democrática para el análisis de las citas recibidas por las revistas. Para estudios posteriores se recomienda la aplicación de otros indicadores bibliométricos que permitan identificar otras tendencias que igualmente ayuden a la mejora de la comunicación científica y consumo de los artículos publicados.

## BIBLIOGRAFÍA